\documentclass[epj]{svjour}
\usepackage{graphics}
\def\ds{\displaystyle}
\begin{document}

\title{Pseudo PT Symmetric Lattice}
\author{C. Yuce 
\thanks{\emph{Present address:} }%
}                     
\offprints{}          
\institute{Physics Department, Anadolu University, Eskisehir, Turkey}
\date{}
%
\abstract{
We study pseudo $\mathcal{P}\mathcal{T}$ symmetry for a tight
binding lattice with a general form of the modulation. Using
high-frequency Floquet method, we show that the critical
non-Hermitian degree for the reality of the spectrum can be
manipulated by varying the parameters of the modulation. We study
the effect of periodical and quasi-periodical nature of the
modulation on the pseudo $\mathcal{P}\mathcal{T}$ symmetry.
 \PACS{
      {11.30.Er,} {Charge conjugation, parity, time reversal, and other discrete symmetries} \and
       {42.82.Et} {Waveguides, couplers, and arrays}\and
      {03.65.-w}{Quantum mechanics}
     } 
}
\maketitle
\section{Introduction}
\label{intro}

Bender and Boettcher showed that the usual requirement of
Hermiticity for the reality of the spectrum for a given
Hamiltonian could be replaced by parity-time
$\mathcal{P}\mathcal{T}$ symmetry \cite{bender}. Because of the
antilinear character of $\mathcal{P}\mathcal{T}$ operator, the
eigenvectors of the Hamiltonian and $\mathcal{P}\mathcal{T}$
operator are not always the same. The spectrum for a
$\mathcal{P}\mathcal{T}$ symmetric non-Hermitian Hamiltonian is
real provided that non-Hermitian degree is below than a critical
number. If it is beyond the critical number, spontaneous
$\mathcal{P}\mathcal{T}$ symmetry breaking occurs. It implies the
eigenfunctions of the Hamiltonian are no longer simultaneous
eigenfunction of $\mathcal{P}\mathcal{T}$ operator and
consequently the energy spectrum becomes either partially or
completely complex. In the past decade, complex Hamiltonian
exhibiting the $\mathcal{P}\mathcal{T}$ symmetry has attracted a
great deal of attention. Recently, $\mathcal{P}\mathcal{T}$
symmetric optical systems with balanced gain and loss has been
experimentally realized \cite{deney1,deney2,deney3}. Not only the
reality of spectrum but also some other physical effects are also
interesting for the $\mathcal{P}\mathcal{T}$ symmetric optical
systems. These are, for example, unconventional beam refraction
and power oscillation \cite{powerosc,refr,ek2}, nonreciprocal
Bloch oscillations \cite{bloch}, unidirectional invisibility
\cite{inv}, an additional type of Fano resonance \cite{fano}, and
chaos
\cite{chaos}. \\
Bendix et al. studied a disordered optical system by considering a
pair of $N$ coupled dimers with two symmetrically placed
impurities \cite{bendix}. They noted that their system is not
$\mathcal{P}\mathcal{T}$ symmetric as a whole (global symmetry),
but it possesses a local $\mathcal{P}_d\mathcal{T}$ symmetry that
admits real spectrum. Recently, the concept of pseudo
$\mathcal{P}\mathcal{T}$ symmetry has been introduced
\cite{enonemli,zhong}. The authors considered a system modulated in such
a way that the corresponding Hamiltonian is neither Hermitian nor
$\mathcal{P}\mathcal{T}$ symmetric. Using the high-frequency
Floquet analysis, they showed that the Hamiltonian admits real
energy spectrum for certain values of parameters in the
Hamiltonian. In \cite{enonemli}, the authors studied periodically
modulated two channel optical coupler with balanced gain and loss.
In this study, we will study pseudo $\mathcal{P}\mathcal{T}$
symmetry for a tight binding lattice with a general form of the
modulation. The tight binding description of $\mathcal{P}
\mathcal{T}$ symmetric lattice has attracted great attention in
recent years
\cite{ek,bendix2,longhiek0,longhiek1,jog0,random2,probConserv,stateTrans,equHerm,random1,disorder1,time1,time2,time3,time4,time6} and is a good model to study pseudo
$\mathcal{P}\mathcal{T}$ symmetry. We consider periodical and
quasi-periodical modulation and study their effects on pseudo
$\mathcal{P}\mathcal{T}$ symmetry.

\section{Model}
Consider an array of $\ds{N}$-site modulated tight binding lattice
with position dependent gain/loss and tunneling amplitude through
which particles are transferred from site to site. The system is
described by the following Hamiltonian in the tight binding regime
\begin{eqnarray}\label{ham}
H= -\sum_{n=1}^N
T_n\left(|n><n+1|+|n+1><n|\right)\nonumber\\+\sum_{n=1}^N\left(f(z)
n+i\gamma_n\right) |n><n|
\end{eqnarray}
where $\ds{T_n}$ is position dependent tunneling amplitude, $f(z)$
is the real valued $z$-dependent potential gradient, and the
complex refractive index related coefficient, $\ds{\gamma_n}$, is
non-Hermitian degree describing the strength of gain/loss material
that is assumed to be balanced, i.e.,
$\ds{\sum_{n=1}^{N}\gamma_n=0}$.  \\
To study pseudo $\mathcal{P}\mathcal{T}$ symmetry, consider the
following general form for the potential gradient
\begin{equation}\label{revf}
f(z)=\omega_0\left( l+\sum_i
\kappa_i\cos(\beta_i{\omega_0}z+\phi_i ) \right)~,
\end{equation}
where $\ds{l{\neq}0}$ is an integer, $\omega_0$ is the modulation
frequency, $\ds{\beta_i}$ are arbitrary real numbers,
$\ds{\phi_i}$ are the initial phases and the constants
$\ds{\omega_0 \kappa_i}$ are the amplitudes of the oscillating
term ($ac$-like term). As can be seen, the potential gradient
function is composed of $dc$-like plus $ac$-like terms.
Apparently, $dc$-like term breaks the $\mathcal{P}\mathcal{T}$
symmetry of the Hamiltonian. So, we expect that the system has
zero threshold for $\mathcal{P}\mathcal{T}$ symmetry breaking,
i.e. $\ds{\gamma_{PT}=0}$. Even in the absence of the $dc$-like term, the $ac$-like term also leads to the
broken $\mathcal{P}\mathcal{T}$ symmetry when the relative phase
between any two component, $\ds{\phi_i-\phi_j}$, is neither zero
nor $\pi$. Below we show that the pseudo $\mathcal{P}\mathcal{T}$
symmetry appears effectively for our non-$\mathcal{P}\mathcal{T}$
symmetric system.\\
Let us investigate the Hamiltonian (\ref{ham}) in more detail. A
common way in the studies of such systems is to find a
$z$-independent effective Hamiltonian. It is well know that the
tunneling parameter is replaced by an effective tunneling
amplitude, $\ds{{T_n^{eff.}} }$ in the high-frequency domain. With
application of the high-frequency Floquet approach, the
Hamiltonian (\ref{ham}) can then effectively be described as
\cite{effective01,effective0,effective2}
\begin{eqnarray}\label{ameffh}
H_{eff.}= -\sum_{n=1}^N
T_n^{eff.}|n><n+1|+{T_n^{eff.}}^{\star}~|n+1><n|\nonumber\\+i\sum_{n=1}^N
\gamma_n |n><n|~~~~
\end{eqnarray}
where star denotes the complex conjugate and the effective
tunneling is given by
\begin{eqnarray}\label{efftun}
\frac{T_n^{eff.}}{T_n}=\overline{ \int_0^z e^{i\eta}dz^{\prime}}
\end{eqnarray}
where overline denotes the average over $z$ and $\eta$ is given by
$\ds{\eta(z)=\int_0^{z}{f(z^{\prime})~dz^{\prime}}}$. Here, it is useful expand the
oscillatory term $\ds{e^{i\eta}}$ in terms of Bessel functions by
using the Jacobi-Anger expansion; $\ds{
e^{i\kappa\sin(x)}=\sum_{m} \mathcal{J}_m ( \kappa )e^{imx} }$,
where $\mathcal{J}_m$ is the
$m$-th order Bessel function of first kind.\\
The $\mathcal{P}\mathcal{T}$ symmetry breaking modulation term in the original Hamiltonian (1) doesn't exist in the effective Hamiltonian. The system is pseudo $\mathcal{P}\mathcal{T}$ symmetric if not the original Hamiltonian but the effective Hamiltonian is $\mathcal{P}\mathcal{T}$ symmetric. Apparently, the effective Hamiltonian is $\mathcal{P}\mathcal{T}$ symmetric if a precise relation between $T_n$ and $\gamma_n$ hold. Below, we will obtain effective
tunneling amplitude for some specific cases and discuss the
physical results.

\subsection{Monochromatic Modulation}

\begin{figure}
\includegraphics{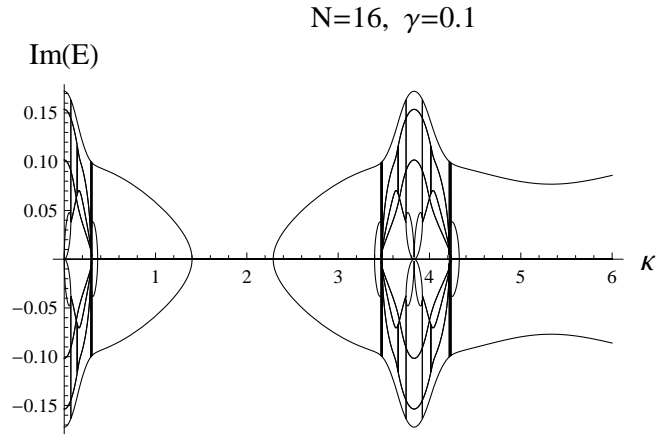}
\caption{The imaginary part of energy eigenvalues for $N=16$,
$l=1$ and $\ds{\gamma=0.1}$ as a function of $\kappa$. The system is not
$\mathcal{P}\mathcal{T}$ symmetric since $l=1$. In the presence of the additional $ac$-like term, $\kappa\neq0$, the system is still in the broken $\mathcal{P}\mathcal{T}$ symmetric phase but the spectrum is real when $\ds{1.4<\kappa<2.3}$. This phase is called pseudo $\mathcal{P}\mathcal{T}$ symmetric phase.  }
\label{ps00bhm}  
\end{figure}

Consider first monochromatic case
\begin{equation}\label{revf241}
f(z)=\omega_0\left( l+ \kappa\cos({\omega_0}z+\phi  ) \right).
\end{equation}
Let us find the corresponding effective tunneling amplitude.
Substitute the Jacobi-Anger expansion into the integration of the equation
(\ref{efftun}). Then we get
\begin{equation}\label{besselacilimi00}
\int_0^ze^{i\eta}dz^{\prime} =\sum_{m} \mathcal{J}_m ( \kappa )e^{im\phi}
S_{m}
\end{equation}
where $\ds{S_{m}=\int_0^z e^{i(m+l)\omega_0{z^{\prime}}}
dz^{\prime} }$. If we take the time average of $\ds{S_{m} }$, i.e.
$\ds{\overline{S_{m}}=\delta_{-l,m} }$, we get the effective
tunneling expression
\begin{eqnarray}\label{son202gh}
\frac{T_n^{eff.}}{T_n}= \mathcal{J}_{-l} (\kappa)e^{-il\phi}
\end{eqnarray}
\begin{figure}
\includegraphics{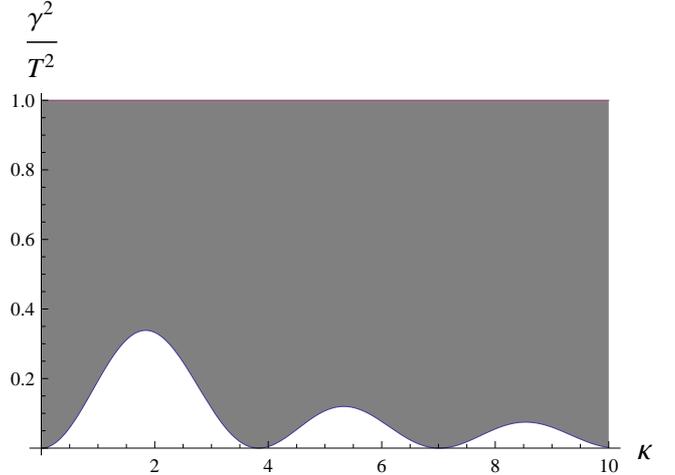}
\caption{The parametric region for the reality of the spectrum for
the parameters $\ds{\gamma^2/T^2}$ and $\ds{\kappa}$ for
$\ds{l=1}$. The shaded (unshaded) region corresponds to parameters
with complex (real) energy eigenvalues. The
$\mathcal{P}\mathcal{T}$ symmetry is broken when $\ds{l}$ changes
from zero to nonzero value. If $\kappa=0$, the spectrum is complex
because of the broken $\mathcal{P}\mathcal{T}$ symmetry. The
presence of $ac$-like term gives rise to the pseudo
$\mathcal{P}\mathcal{T}$ symmetry that guarantees the reality of
the spectrum in a broad range of parameters.}
\label{ps001}  
\end{figure}
We conclude that the presence of the monochromatic modulation
corresponds to a modification of the tunneling amplitude. The
Bessel function $\ds{ \mathcal{J}_{-l} (\kappa)}$ is roughly like
an oscillating sine function that decay proportionally as $\kappa$
increases. In the absence of $ac$-like term, $\kappa=0$, the
Bessel function is always zero, $\ds{ \mathcal{J}_{-l}(0)=0}$.
Therefore, the effective tunneling is suppressed and the spectrum
is complex. This result is a direct consequence of the broken $\mathcal{P}\mathcal{T}$ symmetry that occurs
when $l$ changes from zero to nonzero. The tunneling is partially
restored and the system enters the pseudo $\mathcal{P}\mathcal{T}$
symmetric phase with the additional application of $ac$-like term.
In the pseudo $\mathcal{P}\mathcal{T}$ symmetric region, the
spectrum is real as long as $\ds{\gamma< \gamma_{PT}}$, where the
critical point depends on the effective tunneling amplitude. The critical point becomes zero whenever $\ds{\kappa}$ is a root of Bessel
function of order $l$. So, we say that the pseudo $\mathcal{P}\mathcal{T}$ symmetry is spontaneously broken
at such values of $\ds{\kappa}$.\\
To gain more insight, let us find the energy eigenvalues by
specifying position dependent tunneling amplitude. Analytically
solvable systems are particularly interesting in the study of
non-Hermitian systems. Without loss of generality, suppose that
the system has two sites, $N=2$ (dimer) with two impurities
$\ds{\mp i\gamma}$ and $\ds{T_n=T}$. Therefore, the corresponding
energy eigenvalues are given by
\begin{equation}\label{energyT}
E=\mp\sqrt{|T \mathcal{J}_{-l} (\kappa)| ^2-\gamma^2}.
\end{equation}
Obviously, the parameters $\ds{l}$ and $\ds{\kappa}$ play vital
roles on the reality of the spectrum. The energy eigenvalues are
real in a broad range of the coupling strength $\kappa$ for fixed
$l$. In the fig-{\ref{ps001}}, parametric region for the reality
of the spectrum is shown as a function of the ratio
$\ds{\gamma^2/T^2}$ and $\ds{\kappa}$ at $\ds{l=1}$. The dark and
bright region correspond to complex and real spectrum,
respectively. The boundary between the dark and bright regions
determines the critical point, $\ds{ \gamma_{PT}}$. Note that the
Bessel function is always smaller than one, so the non-Hermitian
degree must satisfy the condition $\ds{\gamma <T}$ for the reality of the spectrum. As can be seen
from the figure, the critical point is zero, $\ds{\gamma_{PT}=0}$,
at $\ds{\kappa=0}$ and hence the energy eigenvalues are complex.
The harmonic modulation enhances the critical point,
$\ds{\gamma_{PT}}$. In other words, the spectrum becomes real unless $\ds{\gamma<\gamma_{PT}}$ although the Hamiltonian is not symmetric under $\mathcal{P}\mathcal{T}$ transformation. This phase is called pseudo $\mathcal{P}\mathcal{T}$ symmetric phase \cite{enonemli}.  \\
Consider another exactly solvable one-dimensional array with
$N=3$, called trimer. Suppose the non-Hermitian degree of the
first and second sites are $\ds{\gamma}$ and $\ds{s\gamma}$,
respectively, where $\ds{s}$ is an arbitrary real valued constant.
Since the gain and loss in the system is balanced, the
non-Hermitian degree of the third one is $\ds{-(1+s)\gamma}$. Suppose $T_1$ and $T_2$ denote the tunneling amplitude between the first and second sites and the second and third sites, respectively. The corresponding energy eigenvalues satisfy the
following equation
\begin{eqnarray}\label{eneif}
E^3-a E-ib =0
\end{eqnarray}
where the coefficients
\begin{eqnarray}\label{ab}
a&=& (T_1^2+T_2^2) \mathcal{J}^2_{-l}(\kappa)-\gamma^2(1+s+s^2)\\
b&=&\gamma \left( \gamma^2 s(1+s)+(T_1^2(1+s)-T_2^2)
\mathcal{J}^2_{-l}(\kappa) \right)
\end{eqnarray}
Equation (\ref{eneif}) allows us to study the reality of the
energy eigenvalues. We demand that the coefficient $\ds{b}$
vanishes (a necessary condition for the reality of the spectrum).
In this case, the three energy eigenvalues are given by
$\ds{E=0}$, $\ds{E=\mp\sqrt{a}}$. Therefore the spectrum is
entirely real if $\ds{a\geq0}$. The energy eigenvalues are degenerate at
$\ds{a=0}$ and symmetric with respect to zero energy value when
$\ds{a>0}$. The necessary condition $\ds{b=0}$ is satisfied if
either one of the following relations is satisfied
\begin{eqnarray}
s=0; T_1=T_2=T\\
\gamma=\gamma_{\mp}=\mp
|\mathcal{J}_{-l}(\kappa)|\sqrt{\frac{T_2^2-T_1^2(1+s)}{s(1+s)}}\label{eneietw}
\end{eqnarray}
where we assume $\ds{T_2^2>T_1^2s(1+s)}$ for the latter relation. In the former case,
$\ds{b}$ vanishes at any $\ds{\gamma}$. However increasing
$\ds{\gamma}$ switches the coefficient $\ds{a}$ to a negative
value at $\ds{\gamma=\gamma_{c}}$ and hence the energy spectrum
becomes complex. The critical point $\ds{\gamma_{c}}$ equals to
$\sqrt{2}T \mathcal{J}_{-l}(\kappa)$ and the energy eigenvalues
are given by
\begin{eqnarray}\label{iiau}
E=0~,~E=\mp \sqrt{2T^2 \mathcal{J}^2_{-l}(\kappa)-\gamma^2}
\end{eqnarray}
In the latter case (\ref{eneietw}), the constant $b$ vanishes if
$\ds{\gamma=\gamma_{\mp}}$. It is interesting to observe that
these are two distinct values that admit real spectrum. The corresponding energy eigenvalues are given by
\begin{eqnarray}\label{eneietw2}
E=0~,~~~~~~~~~~~~~~~~~~~~~~~~~~~~~~~~~~~~~~~~~~~~~~~~~~~~~~~~~~~~~~~~\nonumber\\
E=\mp|\mathcal{J}_{-l}(\kappa)|
\sqrt{T_1^2+T_2^2+\frac{T_1^2(1+s)-T_2^2}{s(1+s)}(1+s+s^2)}~~
\end{eqnarray}
The term in the square root must be greater or equal than zero for
the reality of the spectrum. Comparing the energy expressions
(\ref{iiau},\ref{eneietw2}), we see that the parameters $l$ and
$\kappa$ have nothing to do with the reality of the energy
spectrum for the latter case. Instead, they change the impurity
strength $\ds{\gamma_{\mp}}$ for which the spectrum is real.\\
So far, we have studied the cases with $N=2$ and $N=3$. If the
system has many sites, one can perform numerical computation to
find the energy eigenvalues. We perform numerical calculation when
$N=16$ by supposing the impurities alternate at each site,
$\ds{\gamma_n=(-1)^n \gamma}$ and $T_n=1$. The Fig.(\ref{ps00bhm})
plots the imaginary part of energy eigenvalues at $l=1$ and
$\ds{\gamma=0.1}$. The spectrum is real for certain
range of the parameter $\kappa$ because of the pseudo $\mathcal{P}\mathcal{T}$ symmetric phase.

\subsection{Bichromatic Modulation}

Let us now consider a bichromatic modulation
\begin{equation}\label{revf28}
f(z)=\omega_0\left(l+\kappa_1\cos({{\omega_0}z+\phi_1})+\kappa_2
\cos({\beta{\omega_0}z+\phi_2})\right)
\end{equation}
The bichromatic modulation is periodic if $\ds{\beta}$ is a
rational number and quasiperiodic if it is an irrational number.
Note that $\ds{\beta}$ can be given with a finite number of digits
in a real experiment ($\ds{\beta=p/q}$, where $p$, $q$ are two
coprime positive integers). If $p$ and $q$ are sufficiently large,
then the system becomes effectively quasi-periodic on
the scale of an experiment.\\
Let us find the corresponding effective tunneling amplitude. If we
substitute the Jacobi-Anger expansion into the equation
(\ref{efftun}), we get
\begin{equation}\label{besselacilimijh}
\int_0^ze^{i\eta}dz^{\prime} =\sum_{n,m} \mathcal{J}_n ( \kappa_1
)\mathcal{J}_m ( \frac{\kappa_2}{\beta} ) ~e^{in\phi_1+im\phi_2} ~S_{n,m}
\end{equation}
where $\ds{S_{l,m}=\int_0^z e^{i(l+n+m\beta)\omega_0{z^{\prime}}}
dz^{\prime} }$. Substitution of the time average
$\ds{\overline{S_{n,m}}=\delta_{-l-m\beta,m} }$ into the equation
(\ref{besselacilimijh}) gives the effective tunneling
\begin{eqnarray}\label{son202gh2}
\frac{T_n^{eff.}}{T_n}=e^{-il\phi_1}
\sum_{m}e^{im(\phi_{2}-\beta\phi_1)}\mathcal{J}_{-l-m\beta}
(\kappa_1)\mathcal{J}_{m} (\frac{\kappa_2}{\beta})
\end{eqnarray}
where the phase $\ds{-l\phi_1 }$ affects the effective tunneling
as a whole. There are infinitely many terms in the summation.
However, only a few terms are practically important for small
values of $\kappa$ since the Bessel functions decrease with $m$:
$\ds{\mathcal{J}_{m-1}
(\kappa)/\mathcal{J}_{m}}$.\\
The effective tunneling expressions
(\ref{son202gh},\ref{son202gh2}) show significant difference
between the two cases. Contrary to the monochromatic
modulation, varying the initial phases $\ds{\phi_1}$ and
$\ds{\phi_2}$ change absolute value of the effective tunneling for
the bichromatic case. Therefore, it is possible to observe the
spontaneous pseudo $\mathcal{P}\mathcal{T}$ symmetry breaking not
only by changing $\kappa_1,\kappa_2$ but also by changing the
initial phases at fixed $l$ and $\gamma/T$. Note that the role of
the initial phases depends sensitively on the periodical nature of
bichromatic modulation. If the bichromatic modulation is
quasi-periodic, then only the term $\ds{\mathcal{J}_{-l-m\beta}
(\kappa_1)}$ with $\ds{m=0}$ in the summation (\ref{son202gh2})
survives. Hence, the absolute value of the effective tunneling for
an irrational value of $\beta$ is simplified to
\begin{eqnarray}\label{soafgh2}
\frac{|T_n^{eff.}|}{T_n}= \mathcal{J}_{-l}
(\kappa_1)\mathcal{J}_{0} (\frac{\kappa_2}{\beta})
\end{eqnarray}
We conclude that the effect of the second $ac$-like term with
coupling $\kappa_2$ is the reduced effective tunneling. The
tunneling is lost whenever either $\kappa_1$ or $\kappa_2/\beta$
are the roots of the Bessel function of order $l$ and zero,
respectively. If the bichromatic modulation is periodical, then
the effective tunneling changes with the initial phases. Therefore, the reality of the spectrum
can be controlled by the initial phases $\phi_1$ and $\phi_2$. The effective tunneling
amplitude is not in general real. The complex effective tunneling (\ref{son202gh2}) can be rewritten
as $\ds{T_n^{eff.}=|T_n^{eff.}|~e^{i \Theta}}$, where
$\ds{\Theta}$ is known as  Peierls phase. Let us now study the
effect of a Peierls phase on the energy spectrum. Assume that our
system has periodic boundary conditions. Therefore a momentum
representation is useful to study the system. Suppose that
tunneling amplitude changes with the lattice number $\ds{n}$ such
that $\ds{T_n=T}$ when $n$ is odd and $\ds{T_n=cT}$ when $n$ is
even. The band structure of this system with
$\ds{\gamma_n=(-1)^n\gamma}$ over the Brillouin zone reads
\cite{19,talbot}
\begin{equation}\label{energyT2}
E=\mp\sqrt{\left((c-1)^2+4c
\cos^2(\frac{q-\Theta}{2})\right){|T^{eff.}|}^2-\gamma^2}.
\end{equation}
where the effective tunneling amplitude can be found using
(\ref{son202gh2}). Observe that if the tunneling amplitude is
constant over the whole lattice, i.e. $\ds{c=1}$, then the
spectrum would be complex at any $\ds{\gamma}$ since the cosine
term vanishes at some particular values. As can be seen, the
effect of the Peierls phase is to shift the minimum of the band
structure to a $\ds{q_{min} =\Theta}$. However, the Peierls phase
has nothing to do with the reality of the spectrum and the
critical point, $\ds{\gamma_{PT}=|(c-1)~ {T^{eff.}|}}$.\\
We have seen that the effect of quasi periodical modulation is to reduce the effective tunneling amplitude (\ref{soafgh2}). A question arises. Does the pseudo $\mathcal{P}\mathcal{T}$ symmetric phase appear if the periodic part of the modulation is multicolored? To answer this
question, consider $ f(z)=\omega_0
(l+\sum_{n=1}^{\infty}\kappa_{n} \cos({n{\omega_0}z+\phi_n})
+\kappa \cos({\beta{\omega_0}z+\phi}))$, where the first term is
the Fourier series expansion of a periodic function and $\beta$ is
an irrational number such that $\ds{f(z)}$ is quasi-periodic. One
can study a very large family of potential gradient function using
the Fourier series expansion method. If we use the Jacobi-Anger
expression and evaluate the integral (\ref{efftun}), we get the
absolute value of the effective tunneling amplitude
\begin{eqnarray}\label{ssaef2}
|T_n^{eff}|/T_n=\mathcal{J}_{0}
(\frac{\kappa}{\beta})\prod_{m=1}\mathcal{J}_{-l}
(\frac{\kappa_m}{m})
\end{eqnarray}
where $m$ is the index of multiplication in the product symbol.
Recall that the Bessel function of order $l$ is always less than
one. If there are infinitely many terms in the product, the
effective tunneling is zero independent of $\ds{\kappa_m}$. As a
result, we conclude that the pseudo $\mathcal{P}\mathcal{T}$
symmetric phase disappears if the quasi
periodical modulation is polychromatic.\\
\section{Conclusion}
The idea of pseudo $\mathcal{P}\mathcal{T}$ symmetry has recently been introduced \cite{enonemli}. A pseudo $\mathcal{P}\mathcal{T}$ symmetric Hamiltonian has no $\mathcal{P}\mathcal{T}$ symmetry but can be transformed to a $\mathcal{P}\mathcal{T}$ symmetric Hamiltonian using the high-frequency Floquet approach. In this paper, we have studied pseudo $\mathcal{P}\mathcal{T}$
symmetry for a modulated tight binding lattice with balanced gain
and loss. The Hamiltonian (1), which is neither Hermitian nor $\mathcal{P}\mathcal{T}$ symmetric, has been shown to admit real spectrum in a broad range of parameters. As for a $\mathcal{P}\mathcal{T}$ symmetric Hamiltonian, the spectrum for a pseudo $\mathcal{P}\mathcal{T}$ symmetric is real unless the non-Hermitian degree is below than a critical number. We have shown
that the critical non-Hermitian degree can be manipulated by varying the parameters of the
modulation. We have analytically found the critical number for a dimer and trimer and performed numerical calculation for a lattice with many sites. We have also investigated the effect of periodical and quasi-periodical modulation on the reality of the spectrum.

\end{document}